# Imprinting nanoporous alumina patterns into the magneto-transport of oxide superconductors

J E Villegas[1,*], I. Swiecicki[1], R Bernard[1], A Crassous[1], J Briatico[1], T Wolf[2], N Bergeal[2], J Lesueur[2], C Ulysse[3], G Faini[3], X Hallet[4] and L Piraux[4]

[1]*Unité Mixte de Physique CNRS/Thales, 1 ave. A. Fresnel, 91767 Palaiseau, and Université Paris Sud 11, 91405 Orsay, France*

[2]*LPEM, CNRS-ESPCI, 10 rue Vauquelin, 75231 Paris Cedex 05, France*

[3]*CNRS, Phynano Team, Laboratoire de Photonique et de Nanostructures, route de Nozay, 91460 Marcoussis, France*

[4]*Institut de la Matière Condensée et des Nanosciences, Université catholique de Louvain, Louvain-la-Neuve - 1348, Belgique*

[*] E-mail: javier.villegas@thalesgroup.com

**Abstract.** We used oxygen ion irradiation to transfer the nanoscale pattern of a porous alumina mask into high-$T_C$ superconducting thin films. This causes a nanoscale spatial modulation of superconductivity and strongly affects the magneto-transport below $T_C$, which shows a series of periodic oscillations reminiscent of the Little-Parks effect in superconducting wire networks. This irradiation technique could be extended to other oxide materials in order to induce ordered nanoscale phase segregation.

# 1. Introduction

Modulating the superconducting order parameter at the nanoscale allows manipulating the macroscopic properties of superconductors, which is interesting in view of their technological applications. Examples of this are the possibility of controlling vortex pinning and engineering vortex motion [1], or the realization of planar Josephson junctions [2]. For low-$T_C$ superconductors, nanolithography, etching and lift-off techniques are usually combined to fabricate *ordered* nanostructures in which the film thickness modulation [3], the proximity effect and the presence of a magnetic-field templates [1,4] produce a nanoscale modulation of superconductivity. Those techniques are more difficult to implement and seldom used in oxide high-$T_C$ superconductors. Typically, lithographic features obtained in these materials have sizes above several hundreds of nanometres [5-7], longer than the superconducting characteristic lengths. More common approaches for the introduction of *nanoscale* defects in oxides include heavy ion irradiation [8] or the fabrication of nanocomposite materials [9]. In contrast with lithography techniques, however, the latter produce *disordered* distributions of defects.

In this paper, we report on a non-destructive technique to induce an *ordered* sub-100 nm spatial modulation of the critical temperature in oxide superconducting thin films. This technique utilizes nanoporous alumina templates. These have been used earlier as a substrate for the growth of low-$T_C$ superconductors [10,11], which results in nano-perforated films showing a strong enhancement of the critical currents. This procedure is not feasible for oxide superconductors, as these need to be epitaxially grown and porous alumina is not an appropriate substrate. We propose here a different approach, which uses the alumina template as a mask through which to perform oxygen ion ($O^+$) irradiation. Energetic ions reaching the superconducting film through the pores induce disorder *locally*, which results in the mask nanoscale pattern being "imprinted" into the film. This pattern transfer shows up in the mixed-state magneto-transport, which displays matching effects reminiscent of commensurability between the vortex-lattice and periodic pinning potentials [1,3,5-7], or of Little-Parks oscillations in superconducting wire networks [12,13]. In addition to oxide superconductors, this technique may be well suited to artificially induce *ordered* nanoscale distributions of phases in other oxide materials whose physical properties are strongly dependent upon local disorder.

**2. Experimental details**

For the present experiments, we used two 20 nm thick c-axis $YBa_2Cu_3O_{7-\delta}$ (YBCO) films (A and B) grown on (001) $SrTiO_3$ using pulsed laser deposition (KrF excimer with λ=248nm, energy density 2.4 J·cm⁻² and repetition rate 1 Hz). During growth, the substrate temperature was held at 650°C and the oxygen pressure at 0.355 mbar. The films were cooled down to room temperature in 1013 mb of pure oxygen. For the film A, a protective 5 nm thick amorphous YBCO layer and a 50 nm thick Au layer were subsequently deposited. Optical lithography and ion etching were used to partially remove the Au layer (down to the protective amorphous YBCO layer), and to define Au contact pads.

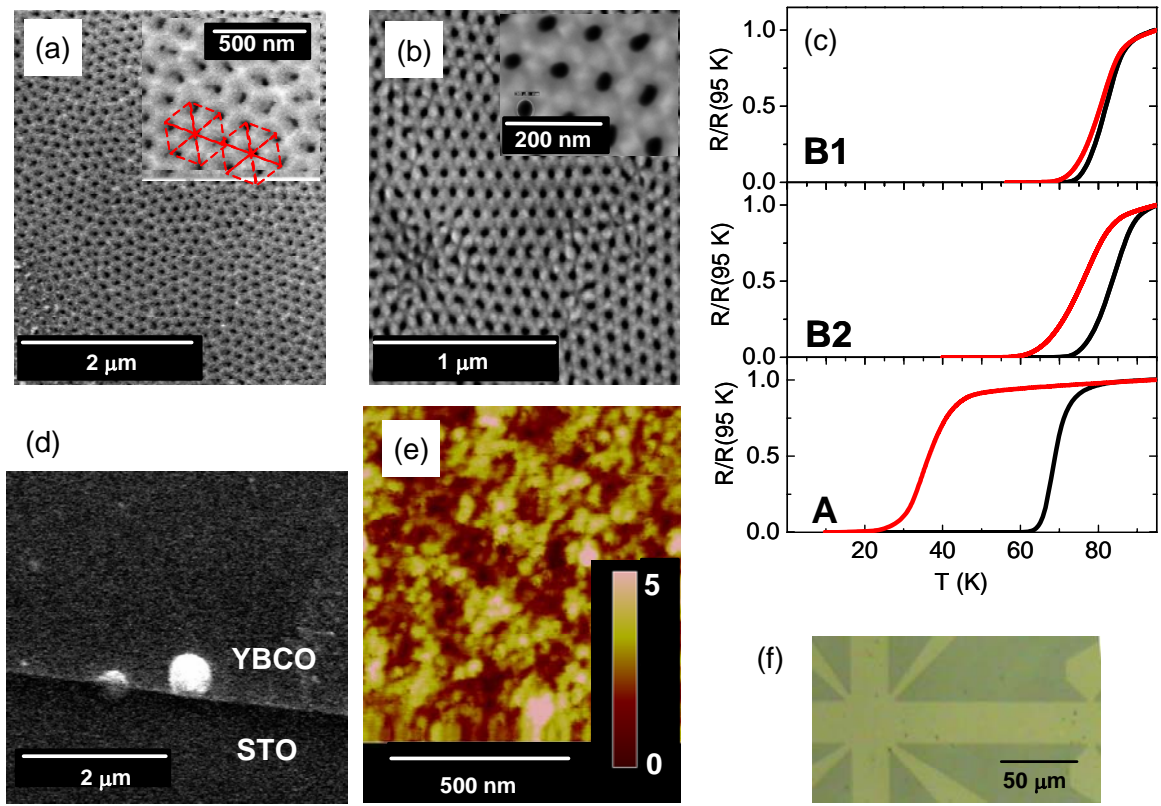

**Figure 1:** (color online) (a) and (b): Scanning electron microscopy (SEM) images of the alumina templates 1 and 2, respectively. The insets show a zoom on the SEM images. The six-fold symmetry of the porous array is highlighted in (a). (c) $R(T)$ (normalized to the resistance at T=95 K) of samples B1 (top), B2 (middle), and A (bottom) prior (black line) and after irradiation (red line). (d) SEM image of sample A after irradiation, nearby one edge of the bridge lithographed for transport experiments. (e) Atomic force microscopy (AFM) image of sample A after ion irradiation. Color grade scale in nanometers. (f) micrograph of the lithographed bridge.

For the film B, the contact pads were defined in subsequent steps: optical lithography, Au sputtering deposition and lift-off of the photo-resist. The film B was diced up in several pieces, two of which (samples B1 and B2) are studied in the present paper. A standard four-probe bridge for resistance measurements (see figure 1 (f)) was optically lithographed and ion etched in all samples. The zero-field resistance *vs*. temperature measurements $R(T)$ of the pristine samples (prior to irradiation) are shown in fig. 1 (c) (black curves). The superconducting critical temperatures (onset of the superconducting transition) $T_C$ and the transition widths $\Delta T_C$ are listed in Table I. Prior to $O^+$ irradiation, a ~60 μm thick porous alumina membrane was fixed on top of the YBCO films by using a few drops of photo-resist. Scanning electron microscopy of the templates (figure 1 (a) and (b)) shows short-range six-fold symmetry. The membrane 1 (pores diameter $\varnothing=55\pm5$ nm and inter-pore distance $d=145\pm10$ nm ) was fixed on sample A. The membrane 2 ($\varnothing=30\pm5$ nm and $d=100\pm5$ nm, figure 1 (a)) was fixed on samples B1 and B2. Oxygen ion ($O^+$) irradiation was made with energies ranging from 110 to 180 keV, total fluence $5\cdot10^{15}$ ions·cm$^{-2}$, and a maximum current of 10 μA to avoid heating. The particular irradiation conditions for each sample are listed in Table I. The projected range of penetration of the 110 keV (180 keV) oxygen ions into alumina and YBCO is about 150 (250) nm [18]. This means that ions are fully stopped by the alumina (thickness ~60 μm) , and do not reach the YBCO film but through the mask pores. Since the track length of oxygen ions is much longer that the YBCO films thickness, ion-induced damage is expected in-depth from the surface to the bottom of the films. *After irradiation the mask was removed by soaking the sample in acetone.* $R(T)$ measurements after irradiation are shown in figure 1 (c) (red curves). The critical temperatures $T_C$ and transition widths $\Delta T_C$ after irradiation are listed in Table 1. In all cases, the critical temperatures were depressed by ion irradiation, and the transition widths enlarged. The effects are more pronounced for sample A, which was irradiated through the alumina template with larger pores ($\varnothing=55\pm5$ nm). Samples B1 and B2 were irradiated through the same template ($\varnothing=30\pm5$ nm) but with different energies, the depression of the critical temperature after irradiation being more pronounced for 110 than for 180 keV. The samples surface was studied using scanning electron (SEM) and atomic force microscopy (AFM). Figure 1 (d) and (e) respectively display SEM and AFM images taken after ion irradiation of sample

A, which show no differences with those taken prior to irradiation (not depicted). The SEM image shows no features other that the eventual presence of nanometric precipitates and "bumps" typical of YBCO thin film growth, which are equally found both before and after irradiation. The root mean square (rms) roughness ~ 1 nm, calculated from AFM images over a 1 μm × 1 μm area, is similar before and after irradiation. We conclude therefore that the irradiation process does not significantly modify the morphology of the sample surface.

| Sample | Before irradiation | | Template | | Irradiation | | After irradiation | | | | |
|---|---|---|---|---|---|---|---|---|---|---|---|
| | $T_C$ (K) | $\Delta T_C$ (K) | $d$ (nm) | $\varnothing$ (nm) | E (keV) | Dose (cm$^{-2}$) | $T_C$ (K) | $\Delta T_C$ (K) | $T_C^{TH}$ (K) | $\mu_0 H_1$ (T) | $\mu_0 H_\phi$ (T) |
| B1 | 88 | 14 | 100±5 | 30±5 | 180 | 5 10$^{15}$ | 87 | 18 | 88 | 0.237±0.004 | 0.24±0.02 |
| B2 | 89 | 16 | 100±5 | 30±5 | 110 | 5 10$^{15}$ | 84 | 24 | 84 | 0.230±0.005 | 0.24±0.02 |
| A | 75 | 12 | 145±10 | 55±5 | 110 | 5 10$^{15}$ | 47 | 21 | 48 | 0.104±0.001 | 0.11±0.2 |

**Table 1 :** Samples parameters : Critical temperature $T_C$ and transition width $\Delta T_C$ prior to irradiation, template inter-pore distance $d$ and pores diameter $\varnothing$, irradiation energy E and dose, critical temperature $T_C$ and transition width $\Delta T_C$ after irradiation, maximum critical temperature expected from simulations $T_C^{TH}$ (see text), experimental matching field $\mu_0 H_1$ and theoretical matching field $\mu_0 H_\phi$ expected from the template inter-pore distance $d$ (see text).

## 3. Results and discussion.

Figure 2 shows the magneto-resistance of sample B2 prior (a) and after irradiation (b), and of sample A after irradiation (c). The magnetic field is applied parallel to the c-axis (perpendicular to the film plane). Note that, for direct comparison of the magneto-transport before and after irradiation, the temperatures for the measurements depicted in (a) and (b) were chosen so that the sample shows the same zero-field resistance in both cases (owing to the different critical temperature and transition widths - see Table 1- this is achieved at slightly different reduced temperatures). All the curves, both for pristine and irradiated films, are symmetric around $H=0$. However, the magneto-resistance is very different after irradiation: the pristine film curve is monotonic and featureless (figure 2 (a)), whereas after irradiation the film (figure 2 (b)) shows a weaker background magneto-resistance and, *most interestingly*, a series of "dips" or inflexion points. These features are also present for the irradiated samples A (figure 2 (c)) and B1 figure 4 (b)). The inflexion points clearly show as minima in the derivative of the magneto-resistance curves $dR/dH$ *vs.* H for the irradiated samples (figures 2 (e)-(f), in contrast with the much smoother derivative observed in the case of the pristine sample (figure 2 (d)).

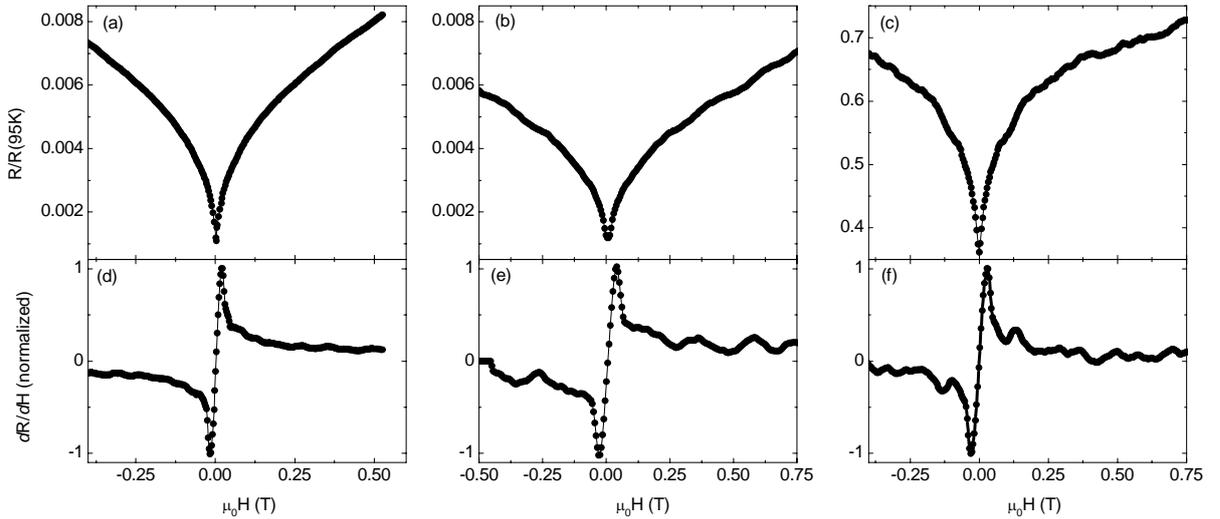

**Figure 2:** Resistance (normalized to the normal-state resistance at T=95 K) as a function of the magnetic field (applied parallel to the c-axis) with injected current $J$=1.25 kA·cm$^{-2}$, for (a) sample B2 prior to ion irradiation, at $T/T_c$=0.74 (b) sample B2 after ion irradiation, at T=0.70$T_c$, and (c) sample A irradiated YBCO film at $T/T_c$=0.74. (d)-(f) Derivatives $dR/dH$ of the curves depicted in (a)-(c), normalized to the maximum slope.

Figures 3 (a) and (b) show the magneto-resistance of samples A and B2 after irradiation, measured at T =0.74$T_C$ and T=0.70$T_c$ respectively, for several injected currents densities, and with the magnetic field applied parallel to the c-axis. The behaviour of both samples is qualitatively similar. The magneto-resistance is ohmic for current levels below J=1.25 kA·cm$^{-2}$, and slightly nonlinear above (it increases by a factor ~1.3-1.5 when the current level is increased one order of magnitude). "Dips" or inflexion points in the magneto-resistance appear at well-defined fields ("matching fields"), which do not depend on the current level. However, the dips are more prominent for the lowest current, and progressively smear out as this is increased. The most pronounced dips appear

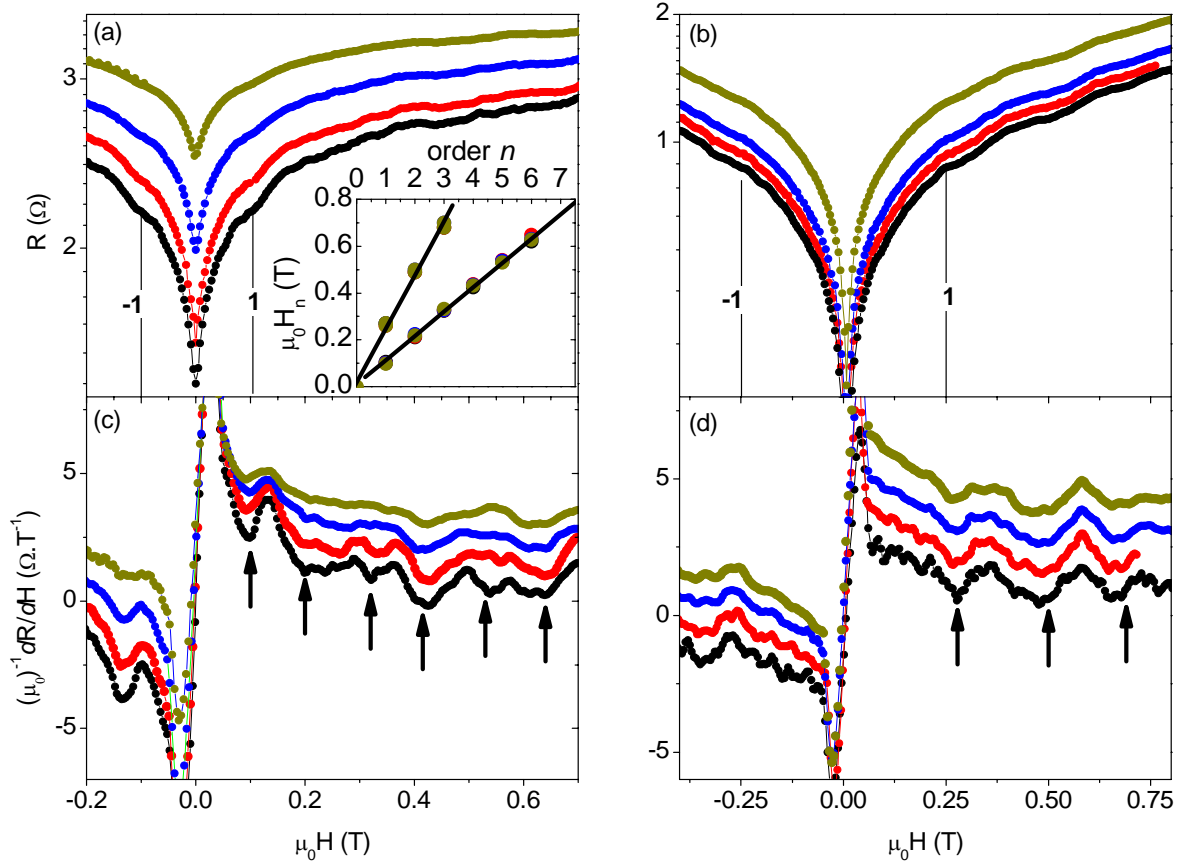

**Figure 3:** (color online) (a) $R(H)$ of sample A at T=0.74$T_C$ with the field applied parallel to the c-axis, for several current densities J=1.25, 2.5, 7.25 and 12.5 kA·cm$^{-2}$ (from bottom to top). The first matching fields (n=1 and n=-1) are indicated by the vertical black lines. Inset: matching fields $H_n$ as a function of the order $n$, for sample A (smaller slope) and B1 (steeper slope). The straight lines correspond to the best linear fits. (b) R(H) for sample B2, at T=0.70$T_C$ and for current densities J=1.25, 3.75, 7.25 and 12.5 kA·cm$^{-2}$ (from bottom to top). (c) and (d): *dR/dH vs.* applied field of the $R(H)$ depicted in (a) and (b), respectively. The derivatives have been vertically shifted for clarity. The arrows indicate integer matching fields.

(symmetrically around $H = 0$) at $\mu_0|H| \sim 0.1$ T and $\mu_0|H| \sim 0.25$ T for samples A and B1, respectively. These matching fields are marked with vertical lines in figure 3 (a) and (b). The series of inflexion points in the magnetoresistance appear as a series of minima in the derivatives $dR/dH$ vs. $H$ (figure 3 (c) and (d)). The series of matching fields $H_n$ as a function of the order $n$ are displayed in the inset of figure 3 (a) for both samples A and B2 The series are well described by $H_n = nH_1$, with $n$ and integer and $\mu_0 H_1 = 0.104$ T (sample A) and $\mu_0 H_1 = 0.23$ T (sample B2) obtained from a linear fit to the experimental data points. These matching fields are in good agreement with the field $\mu_0 H_\phi = (4/3)^{1/2} \cdot \phi_0/d^2$ required to induce a quantum of magnetic flux ($\phi_0 = 2.07 \cdot 10^{-15}$ Wb) per triangular unit cell of the porous alumina template (see Table I for a comparison between $H_1$ and $H_\phi$ for each sample). *Therefore, the matching fields are directly connected with the geometry of the nanoporous alumina template used as a mask for ion irradiation.* The effects described above disappear rapidly as temperature is decreased: the oscillations of the magnetoresistance are observed only within a narrow window of temperatures close to the normal-to-superconducting transition (T/T$_C \sim$33-40 K, T$\sim$60-68 K, T$\sim$54-62 and for samples A, B1, B2 respectively), for which the resistance is ohmic in the low-current level. $R(H)$ progressively becomes featureless and monotonic at lower temperatures, as the current dependence of the resistance becomes non-linear.

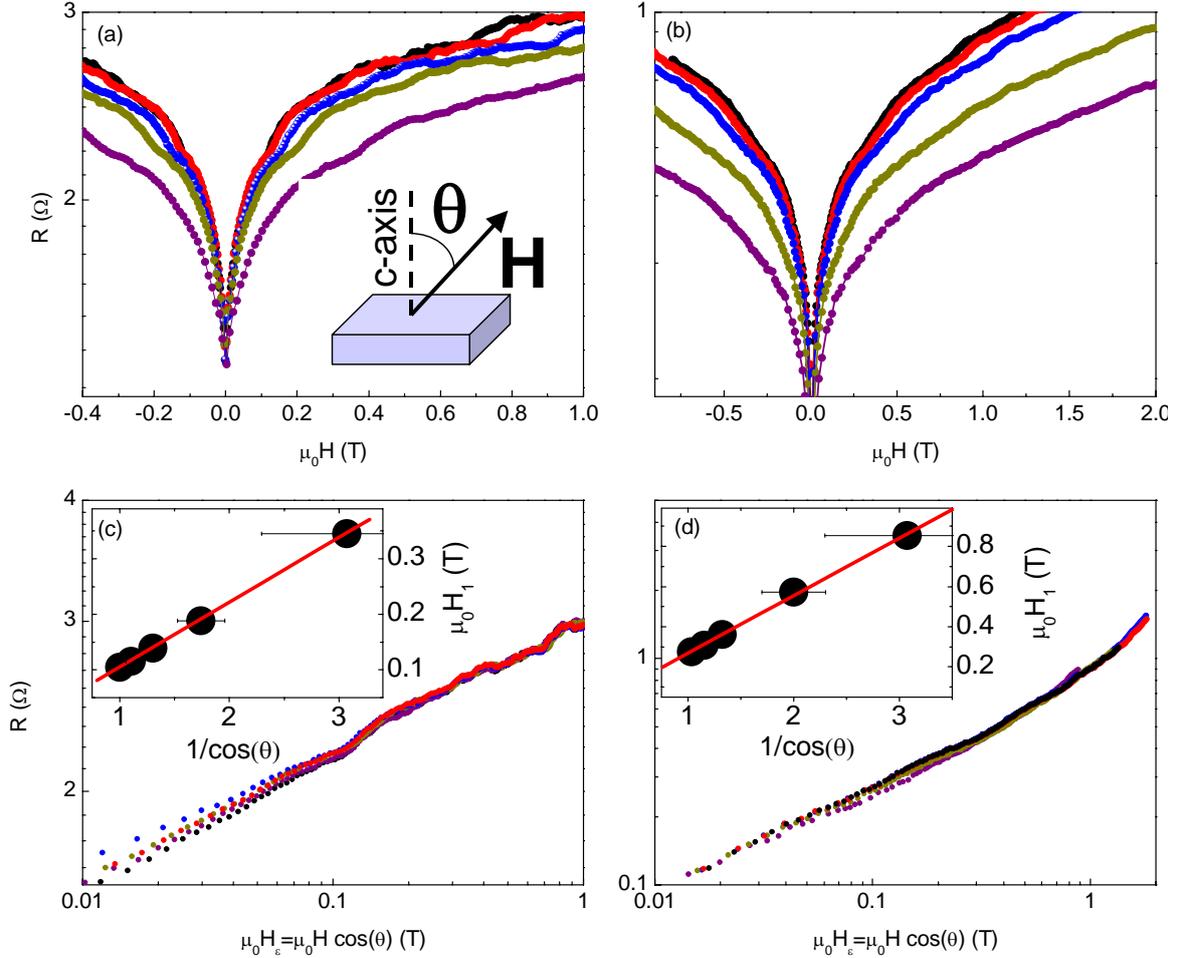

**Figure 4:** (color online) (a) $R(H)$ of sample A at T=0.74$T_C$ and J=1.25 kA·cm$^{-2}$, with the field applied at different angles with respect to the c-axis (see sketch): $\theta$=0, 30, 45, 60 and 75 degrees (top to bottom). (b) $R(H)$ of sample B1 at T=0.75$T_C$ and with J=1.25 kA·cm$^{-2}$ for $\theta$=15, 25, 40, 60 and 70 degrees (top to bottom). (c) and (d) respectively show the same data as is (a) and (b), as a function of the scaled field $H_\varepsilon = H \cdot \cos(\theta)$. Insets: matching field $H_1$ vs. $1/\cos(\theta)$. The straight lines are the best linear fits.

Figures 4 (a) and (b) show the magneto-resistance of samples A and B1, at T=0.74$T_C$ and T=0.75$T_C$ respectively, with J=1.25 kA·cm$^{-2}$, measured for different angles $\theta$ between the c-axis and the applied magnetic field. All samples A, B1 and B2 behave similarly. As the magnetic field is rotated towards the in-plane direction, the background resistance diminishes and the matching fields increase monotonically. In particular, $H_1$ scales as $1/\cos(\theta)$, as shown in the insets of figures 4 (c) and (d). This implies that the matching effects are produced solely by the component of the applied field parallel to the c-axis, *i.e. the matching effects depend on the magnetic flux through the sample surface*. Moreover, all of the $R(H,\theta)$ curves collapse down into a single master curve $R(H_\varepsilon)$ as a function of the effective field $H_\varepsilon = H \cdot \cos(\theta)$ (See figure 4 (b)). This scaling of the background magneto-resistance implies that, in the studied temperature range, the YBCO film is transparent to the in-plane component of the applied field, as in the case of highly anisotropic high-$T_c$ superconductors [14] and ultrathin YBCO films [19].

The experimental results show that matching fields are connected to the geometry of the nanoporous template, which implies that the mask pattern is imprinted into the YBCO thin film. The pattern is transferred as the energetic $O^+$ ions reaching the YBCO film through the pores produce local damage in the form of point defects [15]. These are known to effectively lower the $T_C$ of d-wave superconductors. For doses up to $10^{14}$ ions·cm$^{-2}$, $T_C$ decreases gradually according to an Abrikosov-Gorkov depairing law [16]. Above that level, the system undergoes a metal-insulator transition, and becomes strongly insulating for doses higher than $10^{15}$ ions·cm$^{-2}$ [17]. The typical irradiation angle precision is 1° ($\approx 2 \cdot 10^{-2}$ rad), much larger than the pores aspect ratio ($\sim 10^{-3}$). Therefore, one would expect ions to reach the YBCO film only through a fraction of the pores (those perfectly aligned), and *not at all* through the rest of them. However, this scenario is incompatible with the experimental results: the periodicity of the magnetoresistance oscillations (directly connected to the inter-pores distance *d*) implies that the irradiation is effective through the majority of them. Hence, we conclude that there is a strong ion channelling along the pores, as expected from the large Rutherford scattering cross-section $\sim \sin(\beta)^{-4}$ for grazing incidence ($\beta \leq 1°$) with respect to the pores internal walls [18]. Consequently, we assumed that the film areas directly underneath the pores are irradiated with the nominal dose. Using Monte-Carlo simulations [18] and considering the particular mask geometries, irradiation energies and doses in Table 1, we calculated for each sample the in-plane density of induced point defects (ratio of displaced atoms per existing ones), $\sigma(x, y)$. The contour plots in figure 5 (a) correspond to $\sigma(x, y)$ for samples B1, B2 and A. The spatially varying local critical temperature was calculated from $\sigma(x, y)$ and considering the $T_C$ of the samples prior to irradiation [16]. This is shown in Figure 5 (b), which displays for each sample the local $T_C(x)$ along one of the array axis. The obtained $\sigma(x, y)$ imply that ions spread out when impinging on the YBCO layer. Because of this, point defects are not only induced in the areas directly underneath the membrane pores (circular light-coloured areas in figure 5 (a)), but a non-zero density of defects is induced also in the (inter-pore) areas protected by the alumina mask. The ions spread-out is the smallest and consequently the inter-pore density of defects the lowest for the highest irradiation energy, as one can see by comparing $\sigma(x, y)$ for samples B1 and B2. Consequently, a higher inter-pore $T_C$ is expected for sample B1 that for sample B2 (see figure 5 (b)). This is in good agreement with the experimental observations. Moreover, the onset of the superconducting transition $T_C$ roughly coincides with the maximum critical temperature in the inter-pore areas $T_C^{TH}$ obtained from calculations (see Table 1). On the other hand, a higher inter-pore defects density (1% of that directly beneath the pores) is obtained for sample A than for sample B2, because the membrane pores are larger for the former (~50 nm *vs.* 30 nm), which increases the number of ions spreading into the inter-pore area. This implies a lower $T_C^{TH} \sim 48$ K that for sample B2 (figure 5 (b), lower panel), which agrees with the experimental $T_C \approx 47$ K and the enlarged transition width $\Delta T_C \sim 20$ K (table I). *In summary, both the membrane pattern transfer into the YBCO films, as well as the experimental dependence of their critical temperatures on the membrane parameters, irradiation dose and energy, are explained by the irradiation induced point defects density expected from Monte-Carlo simulations.*

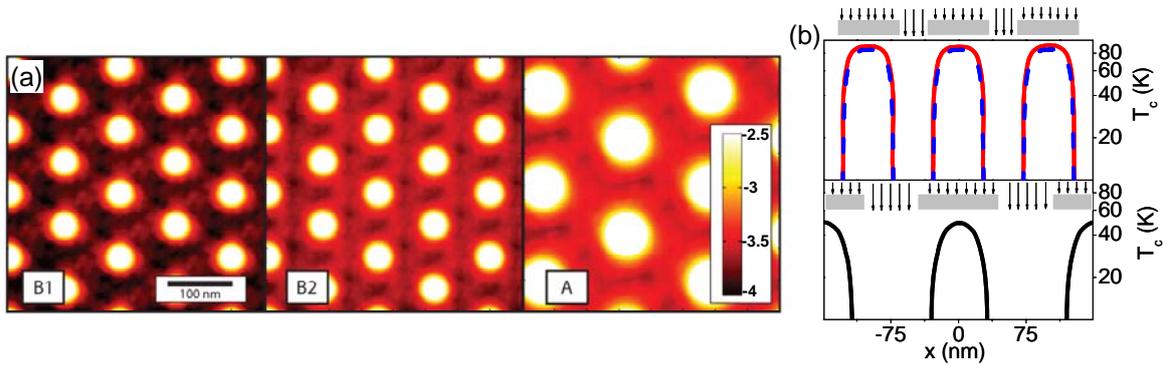

**Figure 5: (a)** Contour plot of the density of defects $\sigma(x,y)$ (see text) induced by irradiation in samples B1, B2 and A. The colour grade scale is logarithmic, and saturates at $\sigma=10^{-2.5}$ (highest) and $\sigma=10^{-4}$ (lowest). The legend indicates the power -4<p<-2.5 for $\sigma=10^p$. **(b)** Calculated local $T_C$ along one of the array axis. The top panel is for samples B1 (red solid) and B2 (blue dashed), and bottom panel for sample A. The grey pads (sketches) represent the alumina mask, and the arrows the impinging ions.

We discuss in what follows the origin of the magnetoresistance oscillations. Matching effects in superconductors are often related to commensurability between the vortex-lattice and the pinning landscape, which results in enhanced vortex pinning. These effects have been profusely studied in low-$T_C$'s [1,3,10,11], and in fewer experiments with high-$T_C$'s [5-7]. Earlier experiments on YBCO films with periodic arrays of holes [7] showed matching effects in the critical current, while here these are observed only in the magnetoresistance well above the irreversibility line. BSCCO single crystals with arrays of micro-holes [6] show matching effects over a much wider region of the vortex-liquid phase than here, and even in the irreversibility line [7]. In all these experiments, the matching effects were related to flux pinning. *The present results differ from these in that here the matching effects appear only above the irreversibility line.* This suggests that matching effects observed here might not be related to flux pinning, but to oscillations of the critical temperature in connection to Little-Parks phenomena [12,13]. In this scenario, the irradiated film would behave as a superconducting wire network, with the wires width equal to the inter-pores distance (~60 to 90 nm). If this length scale is comparable to the superconducting coherence length $\xi(T)$, the critical temperature of the superconducting network will oscillate as a function of the external magnetic field. This comes from the fact that the current in each strand of the network is adjusted in order to maintain the fluxoid quantization in each cell [12]. These effects are expected only close to $T_C$, where $\xi(T)$ diverges, but should rapidly disappear as $\xi(T)$ shortens with decreasing temperature, in agreement with the observed behaviour. The amplitude of the critical temperature oscillations $\delta T_C$ compares well with earlier experiments on YBCO superconducting networks of larger area unit cells (~ $\mu m^2$) [13], in which $\delta T_C$ ~ mK. Since $\delta T_C$ scales with the inverse of the unit cell area [12,13] (~ $10^{-2}$ $\mu m^2$ here), we could expect $\delta T_C$ ~ a few hundreds of mK. This agrees with $\delta T_C \approx 400$ mK obtained from [13] $\delta T_C = (dR/dH)\cdot(dT/dR)\delta H$ (the derivatives of $R(T)$ and $R(H)$ are numerically calculated, and $H = H_1/2 = \delta H$). The observed current dependence [the amplitude of the oscillations decreases as current is increased, figure 3 (b)] is similar to the one observed in earlier experiments [13]. The two-dimensional character of the magnetoresistance is compatible with the wire network picture. Additional evidence to distinguish between the vortex pinning and Little-Park effects scenarios would require to measure the field in-plane magnetoresistance [12], which is not possible here since the film is transparent to that component of the applied field.

## 4. Conclusion

We have shown how to imprint the nanoscale pattern of a nanoporous alumina template in an oxide superconducting thin film by means of $O^+$ ion irradiation. The pattern transfer modifies the magneto-transport, which shows a periodic modulation reminiscent of Little-Parks oscillations in superconducting wire networks. This approach to induce a nanoscale modulation of the physical properties could be extended to other oxide systems where these depend strongly upon local disorder.

## Acknowledgements

Work supported by French Cnano "JN$^2$", ANR "SURF" and "SUPERHYBRIDS-II" and RTRA-Triangle de la Physique "SUPRASPIN" grants, and the Interuniversity Attraction Pole Program (P6/42) - Belgian State - Belgian Science Policy. X. H. acknowledges financial support of the Fund for Training in Research in Industry and Agriculture. J.E.V. thanks Prof. I.K. Schuller for suggesting nanoporous alumina. We thank Y. Le Gall and J.-J. Grob from IneSS at Strasbourg for performing the irradiation.